\documentclass[12pt,a4paper]{article}

\usepackage[latin2]{inputenc}
\usepackage{epic}
\usepackage{latexsym}
\usepackage{epsf}
\usepackage{epsfig}
\usepackage{amssymb,amsmath}

\def\nn{\nonumber}
\def\al{&&\!\!\!\!\!\!\!\!}
\def\d{{\rm d}}

\def\siml{\stackrel{<}{{}_\sim}}

\begin{document}
\thispagestyle{empty}
\begin{flushright}
hep-th/0603014\\
IFT--06--5\\
HD--THEP--06--2                   
\end{flushright}

\vspace*{1cm}

\begin{center}
{\Large\bf Radion stabilization with(out)\\}
\vspace*{5mm}
{\Large\bf Gauss--Bonnet interactions and inflation}
\vspace*{5mm}
\end{center}
\vspace*{5mm} \noindent
\vskip 0.5cm
\centerline{\bf 
D. Konikowska ${}^a$,
M. Olechowski ${}^{a,b}$,
M.G. Schmidt ${}^b$}
\vskip 5mm
\centerline{\em ${{}^{a}}$ Institute of Theoretical Physics, Warsaw
  University}
\centerline{\em ul.\ Ho\.za 69, PL--00--681 Warsaw, Poland}
\vskip 3mm
\centerline{\em ${{}^{b}}$  Institut f\"ur Theoretische Physik,
Universit\"at Heidelberg}
\centerline{\em  Philosophenweg 16, D--69120 Heidelberg, Germany}

\vskip 1cm

\centerline{\bf Abstract}
\vskip 3mm
Radion stabilization is analyzed in 5--dimensional models 
with branes in the presence of Gauss--Bonnet interactions.
The Goldberger--Wise mechanism is considered for static and 
inflating backgrounds. The necessary and sufficient conditions 
for stability are given for the static case. The influence
of the Gauss--Bonnet term on the radion mass and the inter--brane 
distance is analyzed and illustrated by numerical examples.
The interplay between the radion stabilization and the cosmological
constant problem is discussed. 

\newpage
\section{Introduction}

The idea that we live on a brane in a higher dimensional 
space--time has attracted a lot of attention since
Ho\v rava and Witten presented \cite{HoWi}
their 11--dimensional model with 10--dimensional branes, 
motivated by M--theory. Many 5--dimensional models
with 4--dimensional branes (some of them being effective 
models derived from the Ho\v rava--Witten one)
have been discussed since then.
In any such model, the distance between the branes 
should be stabilized. A scalar field 
related to that distance is usually called the 
radion. Several, more or less precise, definitions of 
the radion are used in the literature.  
Of course, not every such a definition  
is reasonable. It was pointed out \cite{ChGrRu} 
that the radion should at least fulfill
appropriate linearized Einstein equations of motion.
It is easy to check that in models where 
only gravity propagates in the bulk the radion 
corresponds to the only gauge--invariant scalar 
perturbation of the metric. Moreover, such a radion 
is massless, which means that the distance between 
branes is not stabilized. Goldberger and Wise 
\cite{GoWi} proposed a class of simple models to 
stabilize this distance. They introduced a bulk 
scalar field with interactions described by some bulk 
and localized brane potentials. Perturbations 
of such a field mix with the scalar perturbations 
of the metric, giving rise to an infinite tower of KK 
states in an effective 4--dimensional theory.
The radion can be defined as the lightest of those 
states. This lowest KK mode might sometimes have
negative mass squared,   
which means that a given configuration is 
unstable. Situations where such
a tachyonic radion may appear were discussed in
several papers \cite{TaMo,CsGrKr,LeSo}. 
In \cite{LeSo} quite general 
criteria for such instabilities were presented.

One of the possible generalizations of this type of 
models takes into account interactions of higher 
order in the curvature tensor. Such interactions 
naturally appear in the $\alpha'$ expansion 
in string theories \cite{strings}. 
The simplest correction is the famous 
Gauss--Bonnet (GB) term. Many aspects of 5--dimensional 
models with GB interactions
have been already discussed in the literature.
However, the influence of GB interactions on the
radion stabilization has not been 
addressed\footnote{
The problem of radion stabilization in 
models with the GB term has been recently addressed
in \cite{AlTr}. However, the definition of the radion 
used in that paper does not seem to be adequate.
It is considered there as the (5,5) component 
of the metric. A potential for such a field is 
introduced  and this 
explicitly breaks symmetries of general relativity.
}. 
It seems that an appropriate way of modeling 
the radion is by introducing an additional 
bulk scalar field \`a la Goldberger and Wise. 
In this work we analyze the radion 
stabilization in models with GB  
interactions and with such a bulk scalar 
field. 
We consider 5--dimensional backgrounds with 
flat static 4--dimensional sections as well as 
with inflating ones. The case with inflation is
especially interesting, because it was shown 
\cite{FrKo} that during inflation stabilization 
of the radion is more difficult. 
We find some generic conditions for radion 
stability in terms of background solutions.
We discuss these conditions using 
analytical arguments as well as results 
of numerical calculations.

Models with scalar fields and Gauss--Bonnet interactions  
have been considered in the literature, but in contexts 
other than radion stabilization.
Problems of singularities and fine tuning in such 
models were discussed in \cite{finetuning}.
Phantom cosmology with GB corrections was 
investigated in \cite{NoOdSa,CaTsSa}. 
In \cite{AmChDa}, the interplay between the 
GB term and the quintessence 
scalar was analyzed. Experimental constraints on the 
quintessence-GB coupling were derived in \cite{Es}. 
Cosmological models with scalar dependent GB 
interactions were considered in \cite{CaNe}. 
In \cite{DeGe}, a scalar field was used to 
support a smooth brane in an Einstein--GB  
gravity model. 
A cosmological model with GB interactions and two scalar 
fields was investigated in \cite{Ne}.

The organization of this paper is as follows: 
In the next section, we present the equations of motion 
and the boundary conditions for models 
with GB interactions and the 
Goldberger--Wise scalar field. The results 
are presented for an ansatz describing the 5--dimensional 
space--time with 4--dimensional inflating de Sitter sections
and two branes.
In section 3, equations of motion and boundary conditions 
for the scalar perturbations in such a model are 
derived and discussed. 
A variational formula for the radion mass is given in 
section 4. It is used later to derive the necessary and 
sufficient conditions for radion stability in the case 
of a vanishing Hubble constant.
In section 5, we compare two classes of solutions present
in models with the GB term. It is shown that the 
stability analysis strongly disfavors the so called "new"
solutions, which appear to be always unstable.
Section 6 contains a discussion of the changes 
caused by the presence of GB interactions in the 
action. The relation 
between the radion stabilization and the cosmological 
constant problem is shortly discussed.
Some results of numerical calculations are presented 
in section 7. Section 8 contains our conclusions.
\section{Background solutions}

We consider 5--dimensional models described by the action
\begin{equation}
S= \int\!\!\d^5x\sqrt{-g}\left[\frac{1}{2\kappa^2}
\left(R + \alpha R^2_{\rm GB}\right)
-\frac{1}{2}(\nabla\Phi)^2-V(\Phi)
-\sum_{i=1}^2\delta(y-y_i)U_i(\Phi)\right]\!,
\label{action}
\end{equation}
where $R^2_{\rm GB}$ is the Gauss--Bonnet term quadratic
in the curvature tensor:
\begin{equation}
R^2_{\rm GB} = R^2 -4R_{\mu\nu}R^{\mu\nu}
+R_{\mu\nu\rho\sigma}R^{\mu\nu\rho\sigma}.
\end{equation}
The scalar field $\Phi$ interacts via the bulk potential
$V(\Phi)$ and two brane potentials $U_i(\Phi)$ located 
at $y_1$ and $y_2$. We do not include extrinsic curvature 
terms in the action, because we work in the up--stairs
picture of the $S^1/{\mathbb Z}_2$ orbifold, where all total 
derivatives integrate to zero. The boundary conditions are obtained
by integrating the equations of motion around the branes. 
This method is equivalent to 
using the Israel junction conditions \cite{Is} after adding 
appropriate extrinsic curvature terms to the action 
\cite{Da}.

We are interested in warped background solutions with
the following ansatz for the metric and the scalar field:
\begin{eqnarray}
&\d s^2 = a(y)^2\left
(-\d t^2 + e^{2Ht}\delta_{ij}\d x^i\d x^j + \d y^2\right),&
\label{ansatz1}
\\[4pt]
&\Phi=\phi(y).&
\label{ansatz2}
\end{eqnarray}
It describes the 5--dimensional space--time warped along  
the $y$--coordinate with the 4--dimensional de Sitter slices  
inflating with the Hubble constant equal to $H$. For $H=0$, 
the same ansatz describes a warped geometry with flat
Minkowski foliation. For simplicity,
the 3--dimensional space has been chosen to be flat. 
Also for simplicity, we assume that inflation is driven by some 
brane fields which do not directly influence the bulk scalar
perturbations, and that the inflaton and radion sectors 
can be treated separately. In this paper, 
the Hubble constant is treated just as a given parameter.
Of course, analysis going beyond 
such a simplification would be very interesting.

The bulk equations of motion for the system described by 
action (\ref{action}) and satisfying ansatz 
(\ref{ansatz1}--\ref{ansatz2}) 
are given by (we use units $\kappa=1$)
\begin{eqnarray}
\label{bulk_bg1}
\phi''+3\frac{a'}{a}\phi'-a^2V'=0\,,
\\[4pt]
\label{bulk_bg2}
\left[\frac{a''}{a}-2\left(\frac{a'}{a}\right)^2+H^2\right]
\frac{\xi}{a^2}+\frac{1}{3}\phi'^2=0\,,
\\[4pt]
\label{bulk_bg3}
3\left[\left(\frac{a'}{a}\right)^2-H^2\right]
\left(1+\frac{\xi}{a^2}\right)-\frac{1}{2}\phi'^2+a^2V=0\,.
\end{eqnarray}
The primes denote differentiations with respect to the 
appropriate (implicit) arguments, 
i.e., with respect to the scalar field $\Phi$ in the 
case of $V$, and with respect to the coordinate $y$ for $a$ 
and $\phi$. To simplify the 
formulae, a new function -- depending on the warp factor, the
Hubble constant and the coefficient of the GB term
--  has been introduced, namely 
\begin{equation}
\xi = a^2 -4\alpha 
\left(\left(\frac{a'}{a}\right)^2-H^2\right)\,.
\label{xi}
\end{equation}
The modifications of the background equations of motion 
caused by the GB term can be writen in a simple way 
in terms of just this single function. 
It is interesting that, as we will show later, this also holds 
for almost all other modifications (equations of motion 
for the perturbations, stability conditions, most of the 
boundary conditions, etc.) considered in this paper. 
The only exception is one of the 
boundary conditions we discuss below.

The boundary conditions are obtained by integration of the 
equations of motion over infinitesimally small intervals 
around the branes. Of course, one must add to the eqs.\  
(\ref{bulk_bg1}--\ref{bulk_bg3}) the contributions coming 
from the brane potentials $U_i$. Such integration of 
eq.\ (\ref{bulk_bg1}) gives the boundary condition for 
the derivative of the scalar background:
\begin{equation}
\lim_{y\to y_i^\pm} 
\frac{\phi'}{a}=\pm \frac12 U'_i.
\label{brane_phi}
\end{equation}
The corresponding condition for the warp factor $a(y)$ 
can be obtained from eq.\ (\ref{bulk_bg2}) 
after using eq.\ 
(\ref{bulk_bg3}) to eliminate $\phi'$. The result reads
\begin{equation}
\lim_{y\to y_i^\pm} 
\left\{
\frac{a'}{a^4} \left[a^2-4\alpha\left(
\frac{1}{3}\left(\frac{a'}{a}\right)^2
-H^2\right)\right]\right\}
=\mp\frac16 U_i
\,.
\label{brane_a}
\end{equation}
The two above equations describe the jumps of the  
${\mathbb Z}_2$ odd functions
$a'(y)$ and $\phi'(y)$ at the ${\mathbb Z}_2$ fixed points
$y_1$ and $y_2$. Observe that the expression in the 
square bracket of the last equation differs from $\xi$
defined in (\ref{xi}) just by a factor 1/3 in front of 
the term containing $a'^2$. The origin of this additional 
factor is the following: 
As the derivative $a'$ is a function discontinuous at
the positions of the branes, the second derivative $a''$ 
contains Dirac delta contributions. Hence  
expressions of the type $a''(a')^2$ must be regularized 
before integration. Any regularization leads to the same result
\begin{equation}
\int_{y_i-\epsilon}^{y_i+\epsilon}a''(a')^2\d y = 
\left.\frac13 (a')^3\right|_{y_i-\epsilon}^{y_i+\epsilon}
\,,
\end{equation}
which gives the factor 1/3 in eq.\ (\ref{brane_a}).

\section{Scalar perturbations}

The main interest in our analysis is the dynamics 
of scalar perturbations around the background metric
given by the solutions to the equations of motion 
(\ref{bulk_bg1}--\ref{bulk_bg3}) with the boundary conditions 
(\ref{brane_phi}--\ref{brane_a}). In general, one could 
perturb all the components of the metric, the scalar field 
and the positions of the branes. Some of such perturbations
correspond to gauge degrees of freedom. This gauge
dependence has been carefully discussed in the literature
\cite{BrDoBrLu,Mukoh,MuKo}. One can choose gauge 
independent combinations of perturbations or work in 
a specific gauge. In the following
we will use the generalized longitudinal gauge.
The scalar perturbations can be written in this gauge as
\begin{equation}
\d s^2 =
a^2\left[\left(1+2F_1\right)
\left(-\d t^2 + e^{2Ht}\delta_{ij}\d x^i\d x^j\right)
+\left(1+2F_2\right)\d y^2\right],
\label{ansatz_F1F2}
\end{equation}
\begin{equation}
\Phi=\phi+F_3,
\label{ansatz_F3}
\end{equation}
where $a$ and $\phi$ are the $y$--dependent background 
solutions to eqs.\ (\ref{bulk_bg1}--\ref{bulk_bg3}) 
and (\ref{brane_phi}--\ref{brane_a}). 
The perturbations $F_i$ are functions of all the 5 coordinates.
This new ansatz has to be substituted into the equations 
of motion obtained from action (\ref{action})
and expanded in powers of the (small) perturbations $F_i$. 
The zeroth order equations are of course given by 
(\ref{bulk_bg1}--\ref{bulk_bg3}) 
and (\ref{brane_phi}--\ref{brane_a}). 
The equations linear in the perturbations can be used to
determine the mass eigenstates for such perturbations in a
given background. Higher order terms of the expansion 
correspond to interactions which are more than bilinear 
in the perturbations. We are mainly interested in the masses 
of the 4--dimensional scalars, especially the radion, 
so we only need the linearized equations.

The off--diagonal components of the linearized Einstein equations
give the following conditions, which must be satisfied in order 
to stay in the longitudinal gauge:
\begin{eqnarray}
\frac{\xi'}{\xi}F_1+\frac{a'}{a}F_2=0,&&
\label{cond_F1F2}
\\
\left(\xi F_1\right)'+\frac13 a^2 \phi' F_3=0.&&
\label{cond_F1F3}
\end{eqnarray}
The diagonal Einstein equations, combined with the background
equations of motion (\ref{bulk_bg1}--\ref{bulk_bg3}), 
give the dynamical equation for the scalar perturbations:
\begin{eqnarray}
\frac{\xi}{a^2}\left[\left(\Box+4H^2\right)F_1 
+ 4\frac{a'}{a}F_1' - 4 \left(\frac{a'}{a}\right)^2F_2\right]
+\frac13 \phi'^2 F_2
\qquad\qquad\qquad\nn\\[4pt]
+ \left(\frac13\phi''+\frac{a'}{a}\phi'\right)F_3 
- \frac13\phi'F_3'=0,
\label{bulk_pert}
\end{eqnarray}
where $\Box$ is the 4--dimensional D'Alembert operator
on the de Sitter slice of the 5--dimensional space--time
(which will become the ordinary flat space D'Alembert operator
when we later set the Hubble constant $H$ to zero).
The boundary conditions for the scalar perturbations read
\begin{equation}
\lim_{y\to y_i^\pm} 
\left(F_3'-F_2\phi'\right)
=\pm \frac12 {aF_3}U''_i.
\label{brane_F2F3}
\end{equation}

Equations (\ref{cond_F1F2}) and (\ref{cond_F1F3}) show that 
in the longitudinal gauge the three scalar perturbations are 
not independent. These equations can be used to eliminate $F_3$ 
and $F_2$, and to rewrite the dynamical equation of motion 
(\ref{bulk_pert}) in terms of $F_1$ alone. In addition, 
taking into account the symmetry of the warped background 
(\ref{ansatz1}--\ref{ansatz2}), we can separate the 
variables in the equation of motion for $F_1$. Defining
\begin{equation}
F_1(t,\vec x,y)=\sum_{m^2}F_{m^2}(y)
\left[\int\d^3 k f_{(m^2,k)}(t)e^{i \vec k \vec x}\right]
,
\end{equation}
we obtain the following equations
\begin{equation}
\ddot f_{(m^2,k)} +3H\dot f_{(m^2,k)}
+\left(a^{-2Ht}\vec k^2+m^2\right)f_{(m^2,k)}=0,
\label{bulk_f}
\end{equation}
\begin{eqnarray}
F_{m^2}''
\al
+\left[2\frac{\xi'}{\xi}-\frac{a'}{a}
-2\frac{\phi''}{\phi'}\right]F_{m^2}'
\nn\\[4pt]
\al
+\left[\frac{\xi''}{\xi}-\frac{\xi'a'}{\xi a}
-2\frac{\xi'\phi''}{\xi\phi'}
-\frac{a^3\xi'}{3a'\xi^2}(\phi')^2
+m^2+4H^2
\right]F_{m^2}=0,
\label{bulk_F}
\end{eqnarray}
where the separation constant $m^2$ turns out to be the  
mass squared of scalars in the effective 4--dimensional 
description.

Let us now rewrite the boundary conditions (\ref{brane_F2F3})
in terms of $F_{m^2}$. The l.h.s.\ of (\ref{brane_F2F3}) 
describes limits (from below or from above) of some expressions 
smooth in the bulk. As a result we can rewrite those 
expressions using the bulk equations 
(\ref{cond_F1F2}) and (\ref{cond_F1F3}) in order to 
replace $F_2$ and $F_3$ with $F_1$. However, this way the second 
derivative of $F_1$ appears in the boundary conditions. 
It can be eliminated with help of the dynamical bulk 
equation (\ref{bulk_F}). The final result takes the form
\begin{equation}
\pm b_{1(2)}\lim_{y\to y_1^+(y_2^-)}
\left(F_{m^2}'+\frac{\xi'}{\xi}F_{m^2}\right) 
+ \left(m^2+4H^2\right)
\lim_{y\to y_1^+(y_2^-)}\left(F_{m^2}\right)
=0\,,
\label{brane_F}
\end{equation}
where
\begin{equation}
b_1=\lim_{y\to y_1^+}
\left(\frac12 a U_1''+\frac{a'}{a}-\frac{\phi''}{\phi'}
\right),
\qquad
b_2=\lim_{y\to y_2^-}
\left(\frac12 a U_2''-\frac{a'}{a}+\frac{\phi''}{\phi'}
\right).
\label{b}
\end{equation}

The above boundary conditions become particularly simple  
for a new variable $Q_{m^2}=\xi F_{m^2}$
(we omit the subscript $m^2$ to simplify the notation):
\begin{eqnarray}
b_1 Q'(y_1^+)+(m^2+4H^2)Q(y_1^+)=0,&&
\nn\\
-b_2 Q'(y_2^-)+(m^2+4H^2)Q(y_2^-)=0.&&
\label{brane_Q}
\end{eqnarray}
The dynamical equation of motion for $Q$ has also 
a very simple form 
\begin{equation}
-\left(p\, Q'\right)'
+q\, Q
=\lambda p\, Q\,,
\label{bulk_Q}
\end{equation}
where $p=3/(2a\phi'^2)$, $q=(a^2\xi')/(2a'\xi^2)$ and 
$\lambda=(m^2+4H^2)$. This is the standard form of
the Sturm--Liouville differential equation.
The boundary conditions (\ref{brane_Q}) can also be 
written in the usual (see e.g.\ \cite{CoHi}) form
\begin{equation}
\frac{\partial Q}{\partial n}(y_i)+\sigma(y_i)Q(y_i)=0,
\label{brane_Q2}
\end{equation}
where in our case $\sigma(y_i)=-\lambda/b_i$.
The $\partial/\partial n$ differentiation is in 
the direction of the outer normal at the boundary. 
For our 1--dimensional equation it is $(-\d/\d y)$ 
at $y_1$ and $(+\d/\d y)$ at $y_2$. The above boundary 
conditions are self--adjoint, but unfortunately quite
unusual, because $\sigma$ depends on the eigenvalue 
$\lambda$. This dependence is caused by the procedure
of obtaining the boundary conditions described before 
eq.\ (\ref{brane_F}).

With such unconventional boundary conditions 
(\ref{brane_Q2}) depending 
on eigenvalues, it is not possible to apply directly 
the analysis of the Sturm--Liouville systems from  
the standard textbooks \cite{CoHi}.
However, it is possible to modify the standard 
arguments to obtain interesting results for 
the case at hand.

\section{Stability conditions}

For the problem of the inter--brane distance stabilization,
the most interesting feature of the analyzed Sturm--Liouville
equation is its lowest eigenvalue $\lambda_0=m_0^2+4H^2$. There are
three possibilities. First: the lowest eigenvalue corresponds 
to positive $m_0^2$, which means that the background
solution is stable against scalar perturbations, and the positions
of the branes are really stabilized. Second: the lowest eigenvalue 
gives vanishing $m_0^2$, the radion is massless and
the inter--brane distance is not stabilized. Third: there is 
at least one tachyon signaling the instability of a given background 
configuration.

We start our analysis with identifying a variational problem 
which is equivalent to the differential equation 
(\ref{bulk_Q}) with the boundary conditions 
(\ref{brane_Q2}). Multiplying eq.\ (\ref{bulk_Q})
by Q and integrating over the 5--th coordinate, one gets
\begin{eqnarray}
\lambda\int p Q_\lambda^2
=\al
\int\left[-(pQ'_\lambda)'Q_\lambda+qQ_\lambda^2\right]
\nn\\[4pt]
=\al
\int\left[pQ_\lambda'^2+qQ_\lambda^2\right]
+\left.p\sigma Q_\lambda^2\right|
=\int\left[pQ_\lambda'^2+qQ_\lambda^2\right]
-\left.\lambda p b^{-1} Q_\lambda^2\right|
,
\qquad
\label{int}
\end{eqnarray}
where the subscript $\lambda$ is added to stress the relation
between an eigenstate and its eigenvalue. 
This equation can be used to express the eigenvalue $\lambda$ 
in terms of the corresponding eigenfunction $Q_\lambda$
as follows\footnote{
In the case of more conventional, 
$\lambda$--independent boundary conditions, the 
contributions from the boundaries appear
with the opposite sign in the numerator \cite{CoHi}.
}: 
\begin{equation}
\lambda=\frac{\int\left[pQ_\lambda'^2+qQ_\lambda^2\right]}
{\int\left[p Q_\lambda^2\right]
+b_1^{-1}\left.\left(pQ_\lambda^2\right)\right|_{y_1}
+b_2^{-1}\left.\left(pQ_\lambda^2\right)\right|_{y_2}
}\,.
\label{lambda}
\end{equation}
The smallest eigenvalue, $\lambda_0$, can be obtained by
minimizing the r.h.s.\ of the above equation over all 
smooth functions $Q$ defined on the interval $[y_1,y_2]$
\begin{equation}
\lambda_0
=\min_Q\left(\frac{\int\left[pQ'^2+qQ^2\right]}
{\int\left[p Q^2\right]
+b_1^{-1}\left.\left(pQ^2\right)\right|_{y_1}
+b_2^{-1}\left.\left(pQ^2\right)\right|_{y_2}}
\right).
\label{lambda0}
\end{equation}

Using the relation between $\lambda$ and $m^2$ one obtains
the following formula for the radion mass
\begin{equation}
m_0^2
=-4H^2+
\min_Q\left(\frac{\int\left[pQ'^2+qQ^2\right]}
{\int\left[p Q^2\right]
+b_1^{-1}\left.\left(pQ^2\right)\right|_{y_1}
+b_2^{-1}\left.\left(pQ^2\right)\right|_{y_2}}
\right).
\label{m0H}
\end{equation}
For a general background (solution to eqs.\  
(\ref{bulk_bg1}--\ref{bulk_bg3})),
the above expression can not be minimized explicitly. 
But we can obtain some bound using just one trial function Q, 
namely the constant one. Recalling the definitions of $p$  
and $q$, we get
\begin{equation}
m_0^2\le
-4H^2
+\frac{\int\d y\left(a^2\xi'/a'\xi^2\right)}
{3\left(\int\d y\left(1/a\phi'^2\right)
+\sum[b_i a(y_i)\phi'^2(y_i)]^{-1}\right)
}\,.
\end{equation}
In the limit of $b_i\to\infty$ (very stiff brane 
potentials, i.e.\ potentials with very large second
derivative) and $\xi\to a^2$ (no GB  
interactions), this expression simplifies to an 
analogous bound obtained in \cite{FrKo}.

In models with inflating branes, i.e.\ for $H^2>0$, stability
of the inter--brane distance occurs when $\lambda_0>4H^2$. 
The formula (\ref{lambda0}) can be used to check this
condition, but in practically all models this can be done
only by numerical calculations. However, some interesting 
analytic results can be obtained for the static 
branes for which $\lambda=m^2$. In such a case, the question 
of stability reduces to the issue of the sign of
$\lambda_0$. In the rest of this section, we assume $H=0$.

It will prove useful to rewrite the differential
equation (\ref{bulk_Q}) in terms of other variables.
Using the definitions 
\begin{equation}
u=\frac{1}{a^{1/2}\phi'}\,Q,
\qquad\qquad
\theta=\frac{a'}{a^{5/2}\phi'}\,,
\label{def_u}
\end{equation}
and the background equations (\ref{bulk_bg1}--\ref{bulk_bg3}), 
the equation of motion (\ref{bulk_Q}) 
can be written in the following simple form
\begin{equation}
u''+\left(\lambda-\frac{\theta''}{\theta}\right)u=0.
\label{bulk_u}
\end{equation}
The differential equations (\ref{bulk_F}), (\ref{bulk_Q}) 
and (\ref{bulk_u}) are not very convenient for background
solutions with $\phi'$ vanishing at some 
$\tilde y\in\,]y_1,y_2[\,$. The reason is that some of 
the coefficients in those equations diverge for $\phi'=0$.
In such cases it is better 
to use a generalization of the Mukhanov variable
\cite{Mukh_var}, defined in our model as
\begin{equation}
v=a^{3/2}\left[F_3-\frac{a}{a'}F_1\right],
\label{def_v}
\end{equation} 
for which the equation of motion reads
\begin{equation}
v''+\left(\lambda-\frac{(1/\theta)''}{(1/\theta)}\right)v=0.
\label{bulk_v}
\end{equation}
Using the background equations (\ref{bulk_bg1}--\ref{bulk_bg3}),
one can show that
\begin{equation}
\frac{(1/\theta)''}{(1/\theta)}
=
\frac{15}{4}\frac{a'^2}{a^2}
-\frac{15}{6}\frac{a^2\phi'^2}{\xi}
-\frac13\frac{a^3\xi'\phi'^2}{\xi^2a'}
+\frac29\frac{a^6\phi'^4}{\xi^2a'^2}
+\frac43\frac{a^5\phi'V'}{\xi a'}
+a^2V''\,,
\end{equation}
which is explicitly regular for vanishing $\phi'$.
The variables $v$ and $Q$ are related by
\begin{equation}
\lambda Q = -a^{1/2}\phi'\theta^{-1}(\theta v)'
\,.
\label{Qv}
\end{equation}

Now we check if there is a massless scalar in the model.
Equation (\ref{bulk_u}) is simple 
enough\footnote{
Unfortunately, for non--zero Hubble constant $H$ the equations 
of motion for the variables $u$ and $v$ can not be written 
in such a simple form as (\ref{bulk_u}) and (\ref{bulk_v}).
}
to be solved for vanishing $\lambda$ almost explicitly:
\begin{equation}
u_0(y) =
\theta(y)
\left[c_1+c_2\int_{y_1}^y\d y'\,
\theta^{-2}(y')\right].
\label{u0}
\end{equation}
Assuming that $\phi'(y)\ne0$ for all $y$ in the bulk, 
one can rewrite (\ref{u0}) in terms of $Q$. 
Using also the boundary condition (\ref{brane_Q}) 
at the first brane (for non--vanishing $b_1$), one 
finds the following result
\begin{eqnarray}
Q_0(y) =\al
\frac{\xi(y)}{\xi(y_1)}
-\frac{a'(y)}{\xi(y_1)a^2(y)}
\int_{y_1}^y\d y'\,\frac{a^2(y')\xi'(y')}{a'(y')}\,,
\label{Q0}\\[4pt]
Q'_0(y) =\al
\frac13[\phi'(y)]^2\frac{a(y)}{\xi(y_1)\xi(y)}
\int_{y_1}^y\d y'\,\frac{a^2(y')\xi'(y')}{a'(y')}\,,
\label{Q'0}
\end{eqnarray}
where the normalization has been chosen as $Q(y_1)=1$.
The boundary condition at the second brane, 
$b_2Q'_0(y_2)=0$, can be fulfilled in several ways.
It seems that one of the possibilities could be the
vanishing of the integral $\int_{y_1}^{y_2}(a^2\xi'/a')$.
This, however, can not be realized because the integrand 
function must be positive for all $y$. The reason is that 
this function is closely related to the kinetic term for 
gravitons. This can be seen by considering (traceless 
and transverse) tensor perturbations $F^{TT}_{\mu\nu}$ 
of the background metric (\ref{ansatz1}). 
By expanding action (\ref{action}) to the second order 
in such perturbations we can obtain  
the appropriate kinetic term. We have found that, 
up to normalization and some total derivatives,
this kinetic term is given by
\begin{equation}
-\frac{a^2\xi'}{a'}\left(\nabla F^{TT}_{\mu\nu}\right)^2\,.
\end{equation}
The coefficient in front of 
$\left(\nabla F^{TT}_{\mu\nu}\right)^2$ must be negative, 
because only then one can define a  
tower of 4--dimensional KK graviton states with 
the standard sign of the kinetic terms
(an effective action for the 4--dimensional KK states
of scalar perturbations was obtained in \cite{KoMaPe}).
Hence for all $y$
\begin{equation}
\frac{\xi'(y)}{a'(y)}>0
\,,
\end{equation}
because otherwise 
some of the graviton KK states become ghost--like.

Let us now return to the discussion of the boundary
condition for solutions given 
by eqs.\ (\ref{Q0}) and (\ref{Q'0}). 
The positivity of the ratio $\xi'/a'$ means that
the integral $\int_{y_1}^{y_2}(a^2\xi'/a')>0$.
Thus the boundary condition at the second brane can be 
fulfilled only when $b_2$ or $\phi'(y_2)$ vanishes.

One can repeat the above reasoning starting from the 
boundary condition at the brane located at $y_2$. 
Choosing the integration constants in (\ref{u0}) in 
an appropriate way (for non--vanishing $b_2$), 
one finds that
the massless mode exists in such a situation only when 
$b_1$ or $\phi'(y_1)$ is zero. Putting both cases together, 
we find that for $\phi'$ non--zero everywhere in the bulk 
the necessary and sufficient condition for existence of 
a massless mode is
\begin{equation}
b_1 b_2 \phi'(y_1) \phi'(y_2)=0.
\end{equation}

Now we identify conditions sufficient for the stability
of the considered warped space--time with branes.
From expression (\ref{lambda}) it is obvious 
that all the eigenvalues are positive if 
$p(y)$, $q(y)$ and $b_i$ are positive. 
This conclusion is correct when the functions 
$p(y)$ and $q(y)$ are finite for all $y$, because
otherwise our Sturm--Liouville system becomes singular,
and a more careful treatment is necessary 
(quantities $b_i$ can be divergent, in
the limit $b_i\to\infty$ the boundary conditions
(\ref{brane_Q2}) reduce to $Q'(y_i)=0$).
The function $p(y)$ diverges at the points where $\phi'(y)=0$.
Using the definition of $\xi$, we get 
\begin{equation}
\frac{\xi'}{a'}=
2\frac{\xi}{a}
+\alpha\frac{4a}{3\xi}(\phi')^2
\,,
\label{xi'a'}
\end{equation}
which shows that $q(y)$ diverges at the points where 
$\xi(y)=0$. The finiteness requirement for 
$p(y)$ and $q(y)$ is equivalent to the requirement 
that $\phi'$ and $\xi$ are non--zero for all $y$.
However, from the background equation of motion 
(\ref{bulk_bg2}) we see that $\phi'$ must be zero
at the points where $\xi$ is zero. Hence the
condition $p(y)\ne\pm\infty$ is stronger than the condition 
$q(y)\ne\pm\infty$.

Thus the brane system is stable if, for all $y$,
\begin{equation}
b_i>0\,,
\qquad
\frac{\xi'(y)}{a'(y)}>0\,,
\qquad
\phi'(y)\ne0\,.
\label{stable}
\end{equation}
These are the sufficient conditions for the stability, 
but we show below that they are in fact also 
the necessary ones.

One of the above conditions, namely the positivity of 
$\xi'(y)/a'(y)$, we have already discussed. 
Any non--positive value of this ratio leads 
to some ghost--like states in the KK tower of the 
4--dimensional gravitons. As a result, the second condition 
in eq.\ (\ref{stable}) is anyway necessary for the stability 
of the model.

Below we will use the following obvious principle 
of variational calculus. Let us consider two problems of 
minimalization of some functionals. If for every function 
the functional in the first problem is not bigger than 
the functional in the second problem then the first minimum 
can not be bigger than the second one.

Let us consider the following two problems.
Both consist in minimizing the 
expression given in (\ref{lambda0}) with the same positive 
$p(y)$ and $q(y)$ but with different $b_i$. 
Let the values of $b_1$ and/or $b_2$ be bigger in the second
problem. Bigger $b_i$ means smaller (at least not bigger) 
denominator in (\ref{lambda0}) and thus bigger 
(at least not smaller) the whole expression to be 
minimized. Applying the above mentioned rule of the 
variational calculus, we see that the lowest 
eigenvalue in the problem with bigger $b_i$ is bigger 
(at least not smaller) than the lowest eigenvalue 
in the problem with smaller $b_i$. The lowest eigenvalue,
$\lambda_0$,  
is a monotonic, non--decreasing function of $b_i$.

We apply the above reasoning to negative $b_1$ or $b_2$.
The monotonic character of $\lambda_0$ as a function 
of $b_i$ means that the lowest eigenvalue for any 
negative $b_1$ ($b_2$)
can not be bigger than the one for vanishing $b_1$ 
($b_2$), which was previously shown to be zero.
We get $\lambda_0\le0$ for $b_1<0$ or $b_2<0$.
On the other hand, $\lambda_0$ can not be equal to zero,
because we have shown that a zero mode exists only if
one of the $b_i$ vanishes (we consider a situation 
with $\phi'(y_i)\ne0$, otherwise there is a zero 
mode for arbitrary $b_i$). A background with a 
negative value of any of the boundary parameters
$b_i$ is unstable.

We have thus shown that the first two conditions in eq.\
(\ref{stable}) are not only sufficient, but also 
necessary for the stability. We will now prove
that the same is true also for the third
condition, which means that all backgrounds with $\phi'$ 
vanishing somewhere in the bulk are unstable. 
Let us assume that $\phi'(\tilde y)=0$ for 
some $y_1<\tilde y<y_2$. One can prove
that in such a case there is at least one 
negative eigenvalue in the problem defined 
by the differential equation (\ref{bulk_Q}) 
with the boundary conditions (\ref{brane_Q2}).

In order to show this, we consider the behavior of 
the solutions to eq.\ (\ref{bulk_Q}) for
large negative $\lambda$. Such behavior can be
found using the variable $v$ defined in (\ref{def_v}).
The differential equation 
for $v$ is regular also for vanishing $\phi'$, 
so the limit $\lambda\to-\infty$ can be 
obtained simply by dropping the $\theta$--dependent 
term in eq.\ (\ref{bulk_v}).
The corresponding solution reads 
\begin{equation}
v_{-\infty}(y)=
c_3\exp[\sqrt{-\lambda}\,(y-y_1)]
+c_4\exp[-\sqrt{-\lambda}\,(y-y_1)]\,.
\end{equation}
Using the relation (\ref{Qv}), we obtain expressions 
for $Q(y)$ and $Q'(y)$ in the limit of large negative 
$\lambda$. The boundary condition (\ref{brane_Q2}) 
at $y_1$ can be fulfilled if $c_3\approx c_4$ 
in the leading order in $1/\sqrt{-\lambda}$. 
Hence $c_3$ is not small compared 
to $c_4$, and away
from the first brane the solution is dominated by 
the exponentially growing term
\begin{equation}
Q_{-\infty}(y)\approx
-\frac{c_3 a^{1/2}(y)\phi'(y)}{\sqrt{-\lambda}}
\exp[\sqrt{-\lambda}\,(y-y_1)]
\label{Q-infty}\,.
\end{equation}
The important feature of this solution is that 
it is proportional to $\phi'(y)$ so it vanishes at 
$\tilde y$ (and changes sign if $\phi'(y)$ changes sign). 
This should be compared to the behavior 
of $Q_0$. As eqs.\ (\ref{Q0}) and (\ref{Q'0}) 
imply that the derivative $Q_0'(y)$ is positive, 
$Q_0(y)$ is also positive for all $y$. 
The solution $Q_\lambda(y)$ changes continuously 
with $\lambda$ changing between 0 and $-\infty$. 
Hence there must exist a negative $\tilde\lambda$ 
for which $Q_{\tilde\lambda}$ has a zero point but 
is nowhere negative. It is easy to see that such
a zero point must be at $y=y_2$, and that 
$Q'_{\tilde\lambda}(y_2)<0$. 
(Both features result from the simple fact that, 
due to the equation of motion, $Q$ and $Q'$ may vanish 
at the same point only for $Q$ vanishing everywhere.)  
The above arguments may be summarized in the following 
form: 
if $\phi'$ vanishes at any point between the branes, 
there is a negative $\tilde\lambda$ for which
\begin{equation}
Q'_0(y_2)>0
\,,\qquad
Q_{\tilde\lambda}(y_2)=0
\,,\qquad
Q'_{\tilde\lambda}(y_2)<0
\,.
\label{lambda_tilde}
\end{equation}
Thus (for non--vanishing $b_2$) 
neither $\lambda=0$ 
nor $\lambda=\tilde\lambda$ is an eigenvalue
of our problem, because in both cases the boundary 
condition at the second brane reduces to 
$Q'(y_2)=0$ and is not fulfilled. It follows 
also that the sign of the l.h.s.\ of the second 
condition in (\ref{brane_Q}) is different for 
$\lambda=0$ and for $\lambda=\tilde\lambda$. 
Thus there must be at least one value of $\lambda$ 
between $\tilde\lambda$ and 0 for which the l.h.s. 
of that equation vanishes. Such negative $\lambda$ 
is an eigenvalue of our problem. Hence  
the radion becomes tachyonic for any background 
solution for which $\phi'$ vanishes anywhere
between the branes. The third condition in 
(\ref{stable}) is necessary for the stability.
This completes the proof that all the conditions 
in (\ref{stable}) are not only sufficient but also
necessary for the stability of our brane model.

This is a generalization of the results obtained in 
\cite{LeSo}\footnote
{In \cite{LeSo}, a different method for analyzing the 
equations of motion for the scalar perturbations  
was used. 
Some features of the phase space associated with
the equations of motion were analyzed to  
find conditions for the lowest eigenvalue. The method used
in the present paper seems to be simpler and easier to
generalize to other models.
}. 
The generalization is twofold. First: we have taken into account 
the possibility of GB interactions present in the 
action. Second: even in the case of the standard Einstein--Hilbert
action we were able to analyze a bigger class of possible 
backgrounds. The authors of \cite{LeSo} assumed that 
$a'(y)\ne0$ for all $y$,  
while we have been able to show that this assumption is in fact 
not necessary.

Let us discuss the last point in some detail. 
The equations of motion for the scalar perturbations 
$Q(y)$ are regular at the points where $a'(y)=0$.
The presence of $a'$ in the denominator in the definition 
of $q(y)$ may be misleading. Equation (\ref{xi'a'}) shows that 
$q(y)=(a^2\xi')/(2a'\xi^2)$ is in fact finite 
for $a'=0$ (at such points $\xi=a^2\ne0$). 
Thus for all backgrounds for which 
$\phi'$ and $\xi$ are non--vanishing for all $y$, 
the Sturm--Liouville system defined by eqs.\ 
(\ref{bulk_Q}) and (\ref{brane_Q2}) is not singular and 
our discussion of the sign of its lowest eigenvalue 
is valid also if $a'$ vanishes at some point(s).

\section{"New" versus "old" solutions}

Action (\ref{action}) contains the Gauss--Bonnet 
term whose strength is parameterized by a constant $\alpha$. 
The results for models with Einstein--Hilbert 
gravity can be obtained from the formulae presented in this  
paper by setting $\alpha=0$ or, equivalently, by replacing 
$\xi$ with $a^2$. However, one should be careful because 
substituting $\alpha$ by zero is not necessarily 
equivalent to the limit $\alpha\to0$. The reason 
is quite obvious: Taking, for example, equation 
(\ref{bulk_bg3}), we see that it is linear 
in the combination $[(a'/a)^2-H^2]$ for $\alpha=0$, 
but is quadratic in this combination for any $\alpha\ne0$. 
For any non--vanishing $\alpha$, there are two solutions 
to this equation (for fixed values of $\phi$ and $\phi'$).
One of them converges to the $\alpha=0$ solutions in the 
$\alpha\to0$ limit. Such type or branch of solutions is 
sometimes called the ``old'' one. 
There are also ``new'' solutions which 
diverge for $\alpha\to0$, but are well defined for any 
$\alpha\ne0$. We will now show that all the solutions 
from this ``new'' branch are strongly disfavored by the stability 
requirements.

Any background solution must fulfill three equations at 
each brane. These are the two boundary conditions 
(\ref{brane_phi}) and (\ref{brane_a}), as well as the 
bulk equation (\ref{bulk_bg3}). Fixing the normalization 
of our metric to be $a(y_1)=1$ and eliminating $\phi'(y_1)$ 
with help of (\ref{brane_phi}), we get the following 
two equations
\begin{eqnarray}
6\left([a'(y_1)]^2-H^2\right)
\left(1-2\alpha\left([a'(y_1)]^2-H^2\right)\right)
=\al
\frac14U'_1(\phi(y_1))-V(\phi(y_1)),
\,\,\,\,\,\,
\label{a'y1a}
\\ 
a'(y_1) \left[1-4\alpha\left(
\frac{1}{3}[a'(y_1)]^2-H^2\right)\right]
=\al
-\frac16 U_1(\phi(y_1))
\label{a'y1b}
\end{eqnarray}
for the two quantities $a'(y_1)$ and $\phi(y_1)$. 
For general potentials $V(\phi)$ and $U_1(\phi)$, this set 
may have several solutions which can be found only numerically. 
However, quite interesting results can be obtained 
analytically in the approximation of a stiff brane potential 
of the form
\begin{equation}
U_1(\phi)=\frac12\mu_1\left(\phi-v_1\right)^2+\ldots
\label{U1}
\end{equation}
with very large $\mu_1$. In such an approximation, the difference 
$\phi(y_1)-v_1$ is small for each solution. This small 
quantity appears linearly in $U'_1$ and quadratically in $U_1$.
Hence in the leading order in $1/\mu_1$ the 
possible values of $a'$ at the brane are given 
approximately by the solutions of
\begin{equation} 
a'(y_1)\left[1-4\alpha
\left(\frac{1}{3}[a'(y_1)]^2-H^2\right)\right]
+\frac16 U_1(v_1)
\approx 0
\,.
\label{a'1}
\end{equation}
The l.h.s.\ of this equation is cubic in $a'(y_1)$
and has two extrema. It is easy to calculate that 
those extrema are at the values of $a'(y_1)$ for which
$\xi(y_1)$, being a quadratic function of $a'(y_1)$,  
vanishes\footnote{
The factor of 1/3 present in (\ref{a'1}) and 
discussed after eq.\ (\ref{brane_a}) is crucial for 
this relation between the extrema of (\ref{a'1}) 
and the sign of $\xi$.
}. 
Thus there is at most 
one solution with positive $\xi(y_1)$. This is 
an "old" solution which reduces to the $\alpha=0$ 
solution in the $\alpha\to0$ limit (if it exists 
for the given parameters). The "new" solutions 
behave like 
$a'(y_1)\sim\alpha^{-1/2}$ in the small $\alpha$ 
limit, and always correspond to negative $\xi(y_1)$.

The above result is very important for the stability 
of the model. By eq.\ (\ref{xi'a'}), the ratio $\xi'/a'$ 
is typically negative for negative $\xi$, which 
leads to an instability of the model. 
Comparing eqs.\ (\ref{xi}) and (\ref{xi'a'}), we see 
that positive $\xi'/a'$ and negative $\xi$ are possible
only for large negative $\alpha$ and large $H$. 
Large values of $\alpha$ are not realistic from the 
phenomenological point of view. Moreover, equation 
(\ref{m0H}) shows that large values of the Hubble constant 
tend to destabilize the model even more.

We have shown above that in the stiff brane potential 
approximation the "new" solutions are unstable. 
Relaxing this approximation is not likely to improve 
the stability of such solutions. It was argued in the 
previous section that the radion mass squared is a 
monotonic function of the parameters $b_i$. Equations
(\ref{b}) show that the $b_i$ decrease when we go away 
from the stiff brane potentials approximation. 
We conclude that all the "new" solutions are  
strongly disfavored by the stability 
conditions.

\section{Role of Gauss--Bonnet interactions}

In the previous section it was shown that the additional 
"new" solutions, which appear after including the GB  
term in the action, are probably all unstable. Only the 
"old" solutions can have a radion with a positive mass squared. 
These "old" solutions converge to the $\alpha=0$ (no GB term) 
solutions in the $\alpha\to0$ limit. This does not, however,  
mean that the addition of GB interactions 
is unimportant. The numerical analysis shows that 
solutions with small but non--zero $\alpha$ can 
substantially differ in some aspects from the solutions 
with $\alpha=0$. The main features of the solutions 
can be understood analytically in the stiff brane potentials 
approximation. For simplicity, we again consider the case with 
a vanishing Hubble constant $H$.

We start with the boundary conditions at the first brane.
As discussed in the previous 
section, the value of $a'(y_1)$ is given by the solution 
to eq.\ (\ref{a'1}). For small $\alpha$, it is
\begin{equation}
a'(y_1)\approx
-\frac16 U_1(v_1)\left[1+\frac{\alpha}{27}U_1^2(v_1)\right]
\,.
\end{equation}
Substituting the above result into eq.\ (\ref{bulk_bg3}), we find
\begin{equation}
\phi'(y_1)\approx
-\frac13 U_1^2(v_1)\left[1+\frac{\alpha}{54}U_1^2(v_1)\right]
+2V(v_1)
\,.
\end{equation}
The last two equations show that the absolute values of 
$a'(y_1)$ and $\phi'(y_1)$ grow with $\alpha$. 
The change of $a$ and $\phi$ with $y$ is faster (slower) for 
positive (negative) $\alpha$ compared to the $\alpha=0$ 
case. One can see that the same is true also away 
from the branes. For sufficiently big $y$ the equations of 
motion (\ref{bulk_bg1}-\ref{bulk_bg3}) are dominated by 
large derivatives and the contributions from the bulk 
potential $V$ become subdominant. In this regime,  
eqs.\ (\ref{bulk_bg1}) and (\ref{bulk_bg3}) give
\begin{eqnarray}
\phi'(y)\approx\al \frac{c}{a(y)^3}
\,,
\label{phi'approx}
\\
[2pt]
\left[a'(y)\right]^2\approx\al
\frac{a^4(y)}{4\alpha}
\left(1\pm \sqrt{1-\frac{2\alpha c^2}{3a^8(y)}}\right).
\label{a'approx}
\end{eqnarray}
Expanding the r.h.s.\ of the last equation in $\alpha$, 
one can see that the background fields change faster 
for positive $\alpha$ and slower for negative $\alpha$.
As the expansion parameter is proportional to $\alpha/a^8$ 
the expansion breaks down for sufficiently small $a$.
The behavior of the r.h.s.\ of (\ref{a'approx}) for small
$a(y)$ depends crucially on the sign of $\alpha$. For negative 
$\alpha$, it tends to a constant, and $a(y)$ can be 
arbitrarily small. For positive $\alpha$, there is a minimal 
value of $a(y)$
\begin{equation}
a^8_{\rm min} \approx \frac23 \alpha c^2
\,,
\label{amin}
\end{equation}
for which $a'$ is still real. It is easy to find the behavior
of different quantities when $a\to a_{\rm min}$, namely:
$a''\to-\infty$, $\xi\to 0$, $\xi'/a'\to+\infty$. 
A solution with positive $\alpha$ ends (if there is no brane 
before, i.e.\ at a smaller value of $y$) at a singularity 
with a finite value of the warp factor. 
The solution can not be extended beyond such point.

The solution for $\alpha=0$ has a different behavior. 
It ends at a singularity for finite $y$ at which 
$a\to0$, $a'\to{\rm const}$, $\phi'\to-\infty$. 
Solutions with negative $\alpha$ also end at 
the singularity $a=0$, and the position of this 
singularity increases with increasing $|\alpha|$.
However, the position of the second brane can not 
be arbitrarily close to such a singularity. The 
reason is that the ratio $\xi'/a'$ becomes negative 
before the position of a singularity, and the model 
is unstable if $\xi'/a'<0$ somewhere between the branes.
The behavior of $\xi'/a'$ can be 
obtained from  the following expansion 
\begin{equation}
\left(\frac{\xi'}{a'}\right)'
=
2a'\left(1-4\alpha\frac{a'^2}{a^4}
-\frac{20}{3}\alpha\frac{\phi'^2}{a^2}
+\frac83\alpha\frac{a}{a'}\phi'V'\right)
+{\cal O}(\alpha^2)
\,.
\label{xi'a''}
\end{equation}
For sufficiently big $y$, this expression is dominated by the 
$\phi'^2$ term which for small $a$ grows approximately as 
$a^{-8}$. It has the same sign as $\alpha$
($a'<0$ because by definition the first brane 
is the positive tension one). Our numerical results 
confirm that indeed $\xi'/a'$ becomes negative before 
a singularity for $\alpha<0$ solutions.
For each background solution, there is a maximal 
possible distance between the branes. For $\alpha\ge0$, 
it is given by the position of a singularity, while 
for $\alpha<0$ it is given by the requirement that 
$\xi'/a'$ must be positive.
The precise value of such a maximal distance depends, 
of course, on the details of the potentials present 
in action (\ref{action}).
This maximal distance between the branes is a decreasing 
function of $\alpha\ge0$. For negative $\alpha$, it is 
more model dependent.

The positivity of $\xi'/a'$ is one of the conditions 
for the stability. Another condition 
is the positivity of the $b_i$ parameters defined 
in eq.\ (\ref{b}). We can easily see how $b_2$ changes 
with $\alpha$. From definition (\ref{b}) and  
equation (\ref{bulk_bg1}) with the $V'$--term neglected,  
one obtains
\begin{equation}
b_2\approx
\frac12aU_2''+4\frac{|a'|}{a}
\,.
\end{equation}
The exact behavior of the r.h.s.\ of this equation 
depends on the details of a model, but one feature
is rather model independent. We have shown that 
the warp factor evolves faster for bigger values of 
$\alpha$. This means that typically $b_2$ increases 
with increasing $\alpha$.

The main consequences 
of adding the Gauss--Bonnet term to the action
are discussed in the two previous paragraphs.
Such a term with a positive coefficient $\alpha$ causes
the inter--brane distance to decrease, and the 
the radion mass squared to increase
(improving the stability of the brane positions). 
The results of adding the GB term with 
a negative coefficient are more model--dependent, but 
in general the stability seems to be worse (the 
radion becomes lighter, and can eventually become
tachyonic).

Let us now discuss the inter--brane distance in more detail.
Of course the second brane can be placed at $y_2$ smaller  
than the maximal value discussed after eq.\ (\ref{amin}). 
A very important 
point is that the boundary conditions (\ref{brane_phi})
and (\ref{brane_a}) must be fulfilled at that second brane.
The values of the l.h.s.\ of both the equations  
are, for a given background solution, known at each $y$.
For a fixed form of the potential $U_2$, also the r.h.s. 
of these equations are known for each $y$. For a general
value of $y$ none of those two boundary conditions 
is fulfilled. The idea of the Goldberger--Wise mechanism 
\cite{GoWi} is that there is a special value of $y$ for 
which both equations (\ref{brane_phi}) and (\ref{brane_a}) 
are fulfilled, and the inter-brane distance is dynamically 
determined to be $y_2-y_1$. 
Of course, in general this can not be achieved. The reason 
is very simple: 
there are two conditions but only one parameter to be 
adjusted, namely $y_2$. So we need a mechanism which 
adjusts the parameters of the Lagrangian in such a way 
that one combination of the boundary conditions 
is fulfilled "automatically". This is just  
another form of the cosmological constant problem. 
Unfortunately, the result given by the Goldberger--Wise 
mechanism depends very strongly on the details of the unknown 
solution to that problem. In the next section 
some results of numerical calculations
are presented to illustrate our analytical
analysis.

\section{Numerical examples}

We consider a simple model similar to that 
discussed in \cite{LeSo}. The bulk and brane 
potentials have the following forms
\begin{equation}
V(\phi)=\Lambda+\frac12 M^2\phi^2
\,,\qquad
U_i(\phi)=\lambda_i+\frac12\mu_i(\phi-v_i)^2
\,.
\end{equation}
We chose the parameters in $V$ and $U_1$ as follows: 
$\Lambda=-15$, $M=1$, $\lambda_1=10$, $\mu_1=100$, 
$v_1=1$ (in $\kappa=1$ units). 
For 5 values of $\alpha$, we solve the boundary condition 
at $y_1=0$ numerically, and integrate the
background equations of motion. Two stability conditions, 
$b_1>0$ and $\phi'(y)\ne0$, are fulfilled in all these
cases. The parameter $\xi'/a'$ as a function of $y$ 
is shown in fig.\ 1. The results are consistent with
our qualitative discussion: away from the first brane,  
$\xi'/a'$ is bigger for bigger values of $\alpha$. 
This effect increases with $y$. For positive $\alpha$,  
the ratio $\xi'/a'$ becomes very large when we 
approach the singularity of the background solution. 
For the Einstein--Hilbert theory ($\alpha=0$), this 
ratio decreases but stays positive up to the singularity. 
For negative $\alpha$, it decreases faster and becomes 
negative before the corresponding singularity.
\begin{figure}[hbt]
\centering\epsfysize=8.0cm \epsfbox{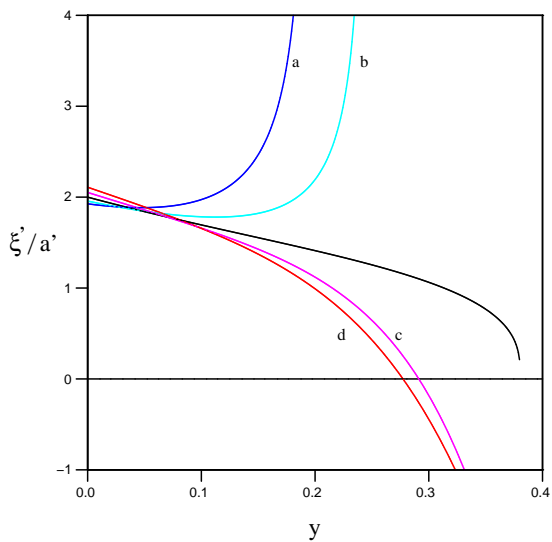} 
\caption[fig1]
{Ratio $\xi'/a'$ as a function of $y$ for different values 
of $\alpha$: 0.01 (a); 0.005 (b); -0.005 (c); -0.01 (d). 
The curve without a label corresponds to a model without the 
Gauss--Bonnet term ($\alpha=0$).
} 
\label{f1}
\end{figure}

Then, for each value of $y$, we solve the boundary conditions 
numerically as if the second brane were positioned 
at that $y$. 
There are two such boundary conditions and three 
parameters in $U_2$ so one of the parameters can 
be fixed. First, we fix $\lambda_2=-\lambda_1$ in order 
to compare our results with those of \cite{LeSo}.
The obtained values of $b_2$  
are plotted in fig.\ 2. The model can be stable only 
for positive $b_2$, so the second brane can be placed only 
at the values of $y$ for which $b_2>0$. 
\begin{figure}[ht]
\centering\epsfysize=8.0cm \epsfbox{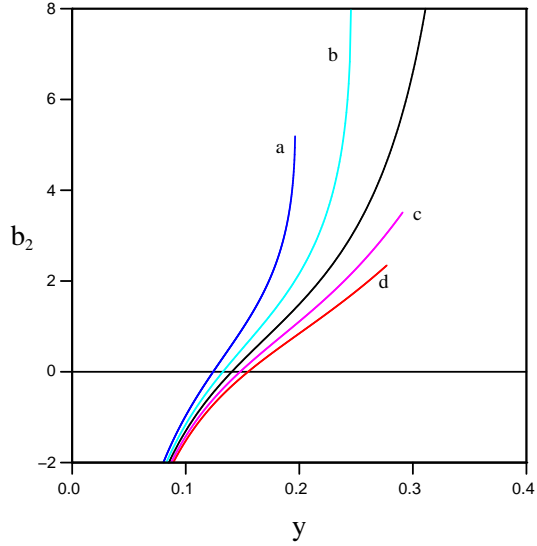} 
\caption[fig2]
{Parameter $b_2$ defined in eq.\ (\ref{b}) as a function 
of the inter--brane distance for $\lambda_2=-\lambda_1$. 
Labels are as in fig.\ \ref{f1}. Curves for $\alpha>0$ 
end at the background solutions singularities, curves 
for $\alpha<0$ end at values of $y$ for which $\xi'/a'$ 
becomes negative.
} 
\label{f2}
\end{figure}

We have discussed the values of $\xi'/a'$ and $b_2$.
In section 4, it has been shown how they are related to the 
radion mass. To illustrate this relation, we have 
calculated the radion mass numerically for some cases. 
For the parameters given above and for 
$y_2-y_1=0.15$, we have found that $m_0^2$ equals  
-0.12, 0.07, 0.40, 1.01, 2.43 
for 5 values of $\alpha$: -0.01, -0.005, 0, 0.005, 0.01,
respectively. Comparing this with figs.\ \ref{f1}
and \ref{f2}, one can see that indeed 
$m_0^2$ grows with the values
of $\xi'/a'$ and $b_2$.
A negative value of $m_0^2$ for $\alpha=-0.01$ reflects 
the fact that in this case $b_2<0$.

We confirm the 
result of \cite{LeSo} that in this particular set--up
there is a minimal inter--brane distance below which 
the model is unstable. With GB interactions,  
this minimal distance decreases (increases) 
for positive (negative) $\alpha$. But there is also an upper
bound on the inter--brane distance. 
For $\alpha\ge0$, it is determined by the singularity 
of the background solution\footnote{
Such an upper bound exists also in the absence 
of the Gauss--Bonnet term, which was not observed in 
ref.\ \cite{LeSo}
} 
while for $\alpha<0$ it is given by value of $y$ for 
which $\xi'/a'$ becomes negative.

One should ask the question: is the above result typical
for the considered model? The answer is: no, it is not. 
To illustrate this, we show in 
figs.\ 3 and 4 the $y$--dependence of 
the parameter $b_2$ for models where the value of 
$\lambda_2$ is changed by 10\%. Such small changes 
in one of the parameters have an important influence on 
the possible inter--brane distance. There is no 
lower bound on $y_2-y_1$ for $\lambda_2=-0.9\lambda_1$ 
or for $\lambda_2=-1.1\lambda_1$. But in the second 
case not all values of $y_2-y_1$ between 0 and the maximal 
value are allowed. There is a window of forbidden values 
of $y$ for which the model becomes unstable while 
being stable outside.

\begin{figure}
\centering\epsfysize=8.0cm \epsfbox{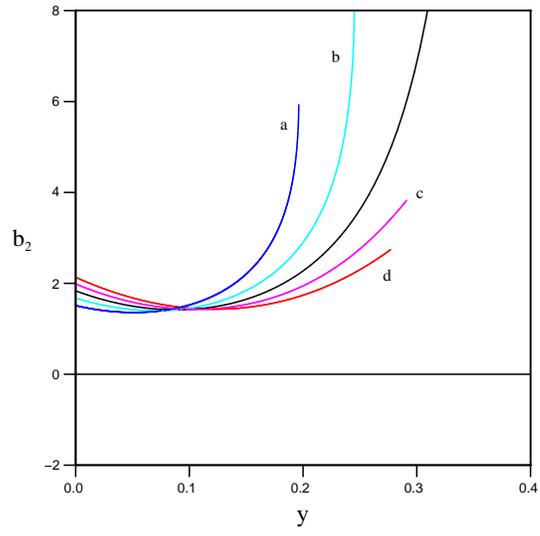} 
\caption[fig3]
{As figure \ref{f2} but for $\lambda_2=-0.9\lambda_1$.
} 
\label{f3}
\end{figure}

\begin{figure}
\centering\epsfysize=8.0cm \epsfbox{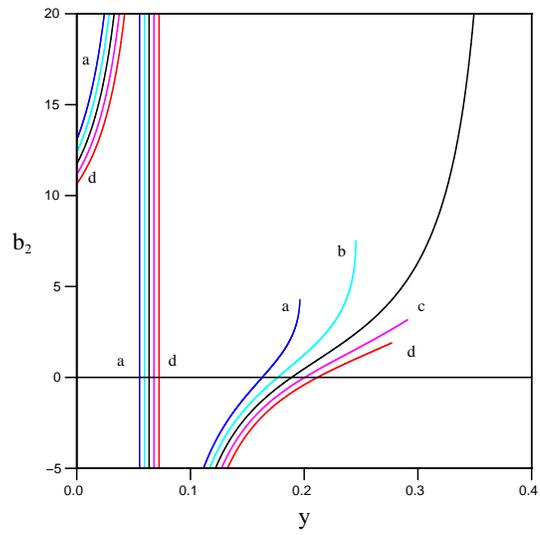} 
\caption[fig4]
{As figure \ref{f2} but for $\lambda_2=-1.1\lambda_1$.
} 
\label{f4}
\end{figure}

In fact, the functional behavior of $b_2$ shown 
in figs.\ \ref{f3} and \ref{f4} is much more 
typical than that in fig.\ \ref{f2}. 
For general values of $\lambda_2$ there 
is no lower bound on $y_2-y_1$. 
For $\lambda_2\siml-\lambda_1$, there are 
two ranges of $y_2-y_1$ for which the model
can be stable. The upper range shrinks when
we decrease $\lambda_2$, and eventually disappears 
for large enough $|\lambda_2|$.
The lower bound on $y_2-y_1$ exists only 
when $\lambda_2\approx-\lambda_1$ up to a few 
percent.
The common feature of all the models is the existence
of an upper bound on the inter--brane distance 
above which the radion can not be stabilized.

The existence of such an upper bound 
can cause problems for model 
building. The most appealing rationale for considering 
the 5--dimensional brane models comes for 
the Ho\v rava--Witten construction motivated by 
M--theory \cite{HoWi}. One of the main features 
of those models is a relatively big length of
the 5--th dimension necessary to obtain a large
enough 4--dimensional Planck scale. Hence one has 
to check whether an upper bound on the inter--brane 
distance in a given model is compatible with 
phenomenological constraints. This is especially 
important after including GB 
interactions, which typically cause the maximal 
inter--brane distance to decrease.

We would like also to stress (again) that radion 
stabilization is quite sensitive to the unknown mechanism 
of solving the cosmological constant problem. In the 
examples presented in this work we fixed $\lambda_2$ and 
used two other parameters in $U_2$ to fix the brane position 
and to set the cosmological constant to zero. 
Of course, we do not know whether these two particular 
parameters have anything to do with the cosmological 
constant problem. Hence one can consider situations 
where parameters other that $\mu_2$ and $v_2$ 
are adjustable. We have checked that, for example, fixing 
$v_2$ gives three classes of solutions 
for $b_2$ similar to those shown in figs.\ 
\ref{f2} to \ref{f4}. However, details 
of a given solution depend on the
actual value chosen for $v_2$.

\section{Conclusions}

We have found the equations of motion for  
scalar perturbations in the presence of the 
Gauss--Bonnet interactions for an inflating 
background. The expression for the mass of the 
lightest of such perturbations, which can be 
identified as the radion, has a form very similar 
to that for the standard Einstein--Hilbert gravity. 
However, the presence of the Gauss--Bonnet term
with a positive coefficient typically increases 
the actual radion mass. This can help to stabilize 
the radion in an inflating background.

The necessary and sufficient conditions for the radion 
stabilization have been found for the case of a 
vanishing Hubble constant. One additional condition 
appears when GB interactions are 
taken into account. The form of the other conditions 
does not change after adding those interactions.
However, the parameters of the model for which those
conditions are fulfilled can change substantially,  
because the character of the background solutions 
changes when the GB term is present in 
the action.

There are two classes of solutions in the presence of 
GB interactions. In the limit where 
the GB coefficient goes to zero, 
solutions from one class just converge to the solutions
in the Einstein--Hilbert theory, while solutions 
from the second class do not. It has been shown that 
the solutions from the second class are strongly 
disfavored by a stability analysis.

Even without inflation, 
numerical calculations are in general necessary 
to check whether the branes can be stabilized 
for a given model. 
The results for the radion stabilization 
are quite sensitive to all (mainly unknown) details 
of models. They depend, for example, on assumptions 
made about the mechanism leading to an acceptably 
small cosmological constant.
This is true for models both with and without the 
Gauss--Bonnet interactions.

However, some typical features of the solutions 
can be found using analytical arguments. 
The most important of them is the
existence of an upper bound on the inter--brane 
distance for each given model. General changes 
caused by the GB term can also be identified. 
Typically, the values of some parameters 
appearing in the formula for the radion mass 
change with $\alpha$ in such a way that this mass 
grows with growing $\alpha$.  
Hence (with all other parameters fixed) stability 
increases if the GB term with a positive 
coefficient is present in the action. At the same time,    
the maximal value of the inter--brane distance becomes 
smaller.

\section*{Acknowledgments}
This work was partially supported by the EU 6th Framework
Program MRTN-CT-2004-503369 ``Quest for Unification''.
M.O.\ was partially supported by 
the DFG Schwerpunktprogramm: ``Stringtheorie in Kontext 
von Teilchenphysik, Quantenfeldtheorie, Quantengravitation, 
Kosmologie und Mathematik'' Schm 561/2-3 
and the Polish MEiN grant 1 P03B 099 29 for years 2005-2007.


\end{document}